%% file: hd180642.tex
\documentclass[useAMS,usenatbib]{mn2e}
\usepackage{graphicx}
\include{ref_abbrevs}
\title[Chaos in large-amplitude pulsators]{Chaos in large-amplitude pulsators: application to the $\beta$\,Cep star HD\,180642}
\author[Degroote, P.]{Degroote, P.$^{1,2}$\thanks{E-mail:
pieterd@ster.kuleuven.be}\\
$^{1}$Instituut voor Sterrenkunde, Celestijnenlaan 200D, 3001 Heverlee, Belgium\\
$^{2}$Stellar Astrophysics Center, Department of Physics and Astronomy, Aarhus University, Ny Munkegade 120, 8000 Aarhus C, Denmark}
\begin{document}

\date{Accepted 2013 February 21.  Received 2013 February 21; in original form 2012 July 12}

\pagerange{\pageref{firstpage}--\pageref{lastpage}} \pubyear{2013}

\maketitle

\label{firstpage}

\begin{abstract}
 The CoRoT observations of the $\beta$\,Cephei star HD\,180642 uncover an unexpectedly rich frequency spectrum, in addition to several heat-driven modes. So far, two processes have been proposed to explain this behaviour: the presence of stochastic oscillations, and the excitation of time-dependent frequencies by nonlinear resonance.
  
   I argue for a third explanation for the observations, in the form of chaos due to the nonlinear behaviour of the dominant radial mode.
  
   The long-term frequency stability of the dominant radial mode is studied using archival data spanning roughly 20 years. Nonlinear time series analysis techniques are applied to the CoRoT observations, and the observations are compared with simulations of a simple nonlinear oscillator.
  
   I show that chaos offers one single explanation for many of the observed features, such as the structure in the autocorrelation of the power spectrum, the long-term frequency shift, the power excess and the wide range of frequencies in the power spectrum. However, the mixture of opacity-driven linear oscillations and nonlinear oscillations complicate the nonlinear time series analysis techniques.
\end{abstract}

\begin{keywords}

\end{keywords}

\section{Introduction}
HD\,180642 is a large-amplitude $\beta$\,Cep star observed with the CoRoT space satellite \citet{fridlund2006}. Its rich frequency spectrum \citep{degroote2009a} immediately led to a debate on the interpretation. \citet{belkacem2009} interpreted the power excess at high frequencies ($130-300$\,$\mu$Hz) as due to solar-like oscillations. \citet{degroote2009a}, however, noted that there was also power excess at lower frequencies, and focused on the nonlinear resonances of the few high-amplitude modes and their effect on the high-frequency regime. These were subsequently confronted with evolutionary models by \citet{aerts2011}. In the latter paper, the left-over frequencies were interpreted as being excited by nonlinear resonance with the dominant, high-amplitude radial mode.

Aside from the high-quality CoRoT observations, archival ground and space based data \citep{aerts2000,uytterhoeven2008} span roughly 20 years, allowing to study long term temporal stability of the detected modes. Period changes are observed in various types of pulsating stars, in various stages of their evolutionary stages. Such changes were discovered in white dwarfs \citep{winget1991_mnras} and $\delta$\,Sct stars \citep{breger1998}. \citet{pigulski2008b} observed a frequency decrease ($D\approx -1.06\times 10^{-6}$\,d$^{-2}$) in the binary $\beta$\,Cep star HD\,168050 using ASAS data, suggesting that period changes might occur in other $\beta$\,Cep stars as well. The large scatter in the O-C diagrams of Cepheids \citep{szabo2011_mnras} hints for variability in the dominant frequency, while \citet{kolenberg2010_mnras} finds smooth modulations of the main frequency in RR\,Lyr stars. Finally, \citet{zijlstra2004} identifies different types of period changes, from sudden changes or irregular patterns, to smooth modulations in Mira stars. Bar the two first examples, most of the observed period changes are larger than expected purely from evolutionary effects, or even have the wrong sign. This can possibly be attributed to `evolutionary weather' (D.~Kurtz, priv. comm.), in which case the evolution of the star on the main sequence is not simply described by a continuous expansion of the star. \citet{breger1998} propose two additional origins of period changes: the existence of a stellar companion and nonlinear mode interactions. The latter explanation is definitely appealing in the case of large amplitude pulsators. Nonlinearity can also give rise to period doubling events \citep[as is observed in RR\,Lyrae star, ][]{kolenberg2010_mnras}, and chaotic behaviour. The latter comes in many different forms, from erratic variability without a clear periodic signature, to mild forms of chaos where a period signature is clearly detected, but not every cycle is exactly reproduced from one to the next. This can introduce a characteristic signature in the Fourier spectrum.

In this paper, I report on the detection of a period change in the $\beta$\,Cep HD\,180642, for which over 20\,years of photometric observations are available, among which a 150\,d continuous, high quality CoRoT light curve. Aside from the apparent period change, also power excess is observed. As a third alternative for the interpretations given by  \citet{belkacem2009,aerts2011}, I show that chaotic behaviour can explain both the period change and power excess. In Section 2, I present the archival photometric data, which is used in Section 3 to derive the magnitude of the frequency shift of the dominant radial mode. Subsequently, the power excess is studied and compared with simulations of a simple nonlinear oscillator. Nonlinear time series analysis techniques are then applied in an attempt to firmly establish the presence of chaos.
\section{Observations}

\begin{figure}
\includegraphics[width=84mm]{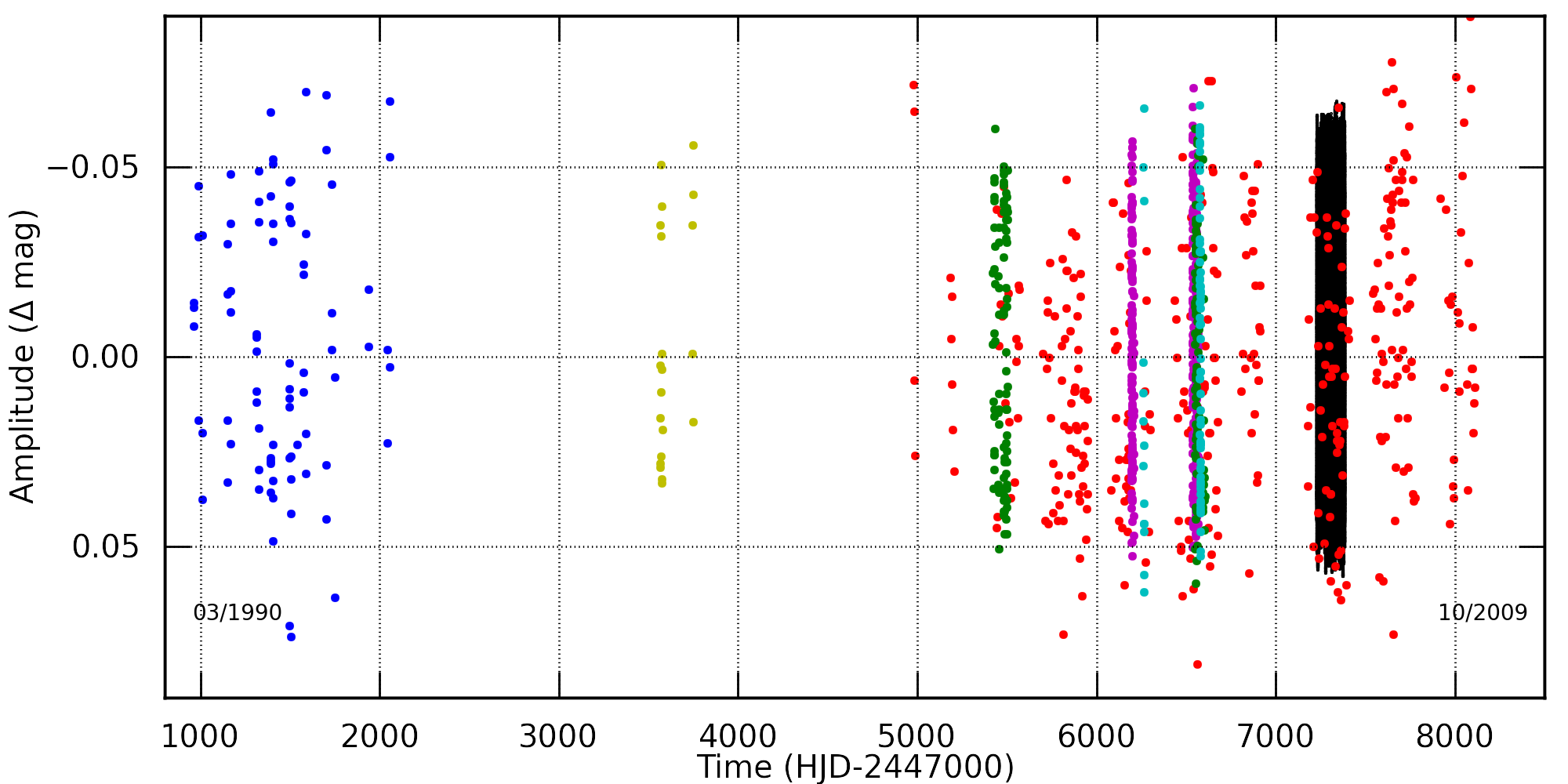}
\caption{Overview of photometric time series of HD\,180642 (blue denotes Hipparcos data, yellow is data from the Swiss 70\,cm telescope, green from Mercator/P7, red from ASAS, cyan from Konkoly, magenta from SPMSNO and black from CoRoT.}\label{fig:hd180642:alldata}              
\end{figure}

The Hipparcos satellite observed HD\,180642 from 12 March 1990 for almost three years. Since then, 191 observations assembled with the photomultiplier P7 attached to the 0.7m Swiss telescope at La Silla and to the 1.2m Mercator telescope at La Palma were made and added to the dataset. Next, the data from two observing campaigns at the Konkoly Observatory in Hungary and at the National Observatory in Sierra San Pedro Martir in Mexico were added. Moreover, there are 379 archival ASAS data points \citep{pigulski2008}. Finally, I add the CoRoT light curve \citep[see ][ for more details]{auvergne2009_mnras,degroote2009b_mnras}, containing more than 300\,000 single observations over a time span of about 150\,d, with far superior precision than the previous datasets. This brings the total time span of observations to almost 20 years (Fig.\,\ref{fig:hd180642:alldata}). The characteristics of the different datasets (year of first measurement $t_0$, total time span $T$ and number of observations $n$), as well as the first frequency value $f$ determined for each of these data sets are summarized in Table\,\ref{tbl:hd180642:ground_based} and shown in Fig.\,\ref{fig:hd180642:freqsummary} (top left panel). The frequency from the Konkoly observatory deviates strongly from the others, nearly simultaneous data sets. The reason is the awkward window function of this dataset, which leads to an alias of the true frequency having the highest amplitude in the power spectrum.

\begin{table}
\caption{Archival photometric datasets used in this paper.}

\centering\begin{tabular}{lcrrll}\hline
          & $t_0$ (y) & $T$~(d) & $n$ & $f$ (d$^{-1}$) \\\hline\hline
Hipparcos$^{(1)}$     & 1990 & 1091 & 83       & 5.48710(4)  \\
Swiss 70cm/P7$^{(2)}$ & 1997 & 187  & 20       & 5.4868(3)   \\
ASAS$^{(3)}$          & 2001 & 2798 & 379      & 5.486912(6) \\
Mercator/P7$^{(4)}$   & 2002 & 1184 & 171      & 5.48693(2)  \\
Konkoly$^{(4)}$       & 2004 & 319  & 83       & 5.48734(5)  \\
SPMO/SNO$^{(4)}$        & 2004 & 371  & 335      & 5.48697(7)  \\
CoRoT/SISMO$^{(5)}$   & 2007 & 156  & 379\,785 & 5.48688(1)  \\ \hline
All           & 1990 & 7148 & 380\,787 & 5.4869441(3)\\ \hline\end{tabular}
\label{tbl:hd180642:ground_based}
\footnotesize{(1) \citet{aerts2000} (2) \citet{briquet2009_mnras} (3) \citet{pigulski2008} (4) \citet{uytterhoeven2008} (5) \citet{degroote2009a}}
\end{table}

\section{Data analysis}

\subsection{Period change}

The evolution of the dominant period has already been explored briefly by \citep{degroote2009b_mnras}, but their analysis was not quantitative and no physical interpretation was given.

\begin{figure}
\centering

\includegraphics[width=0.49\columnwidth]{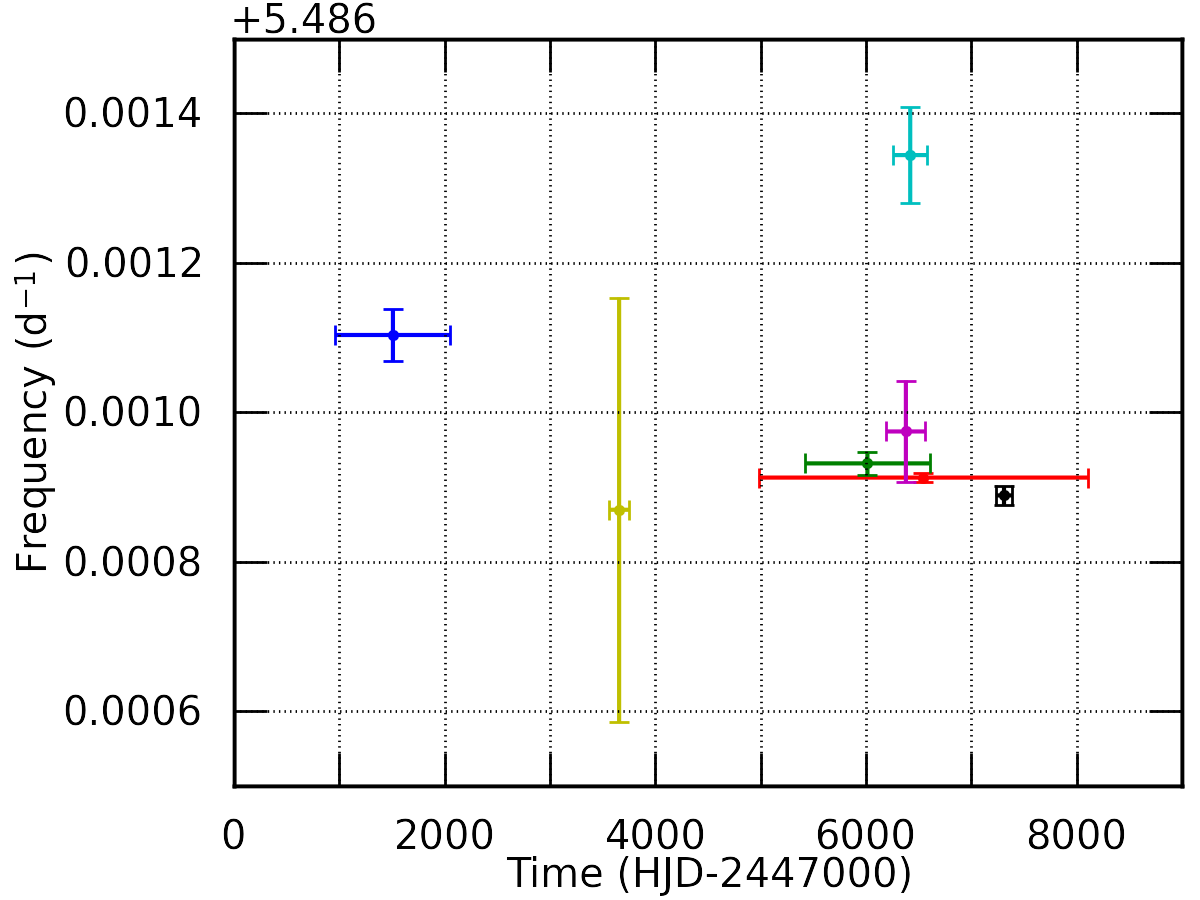}
\includegraphics[width=0.49\columnwidth]{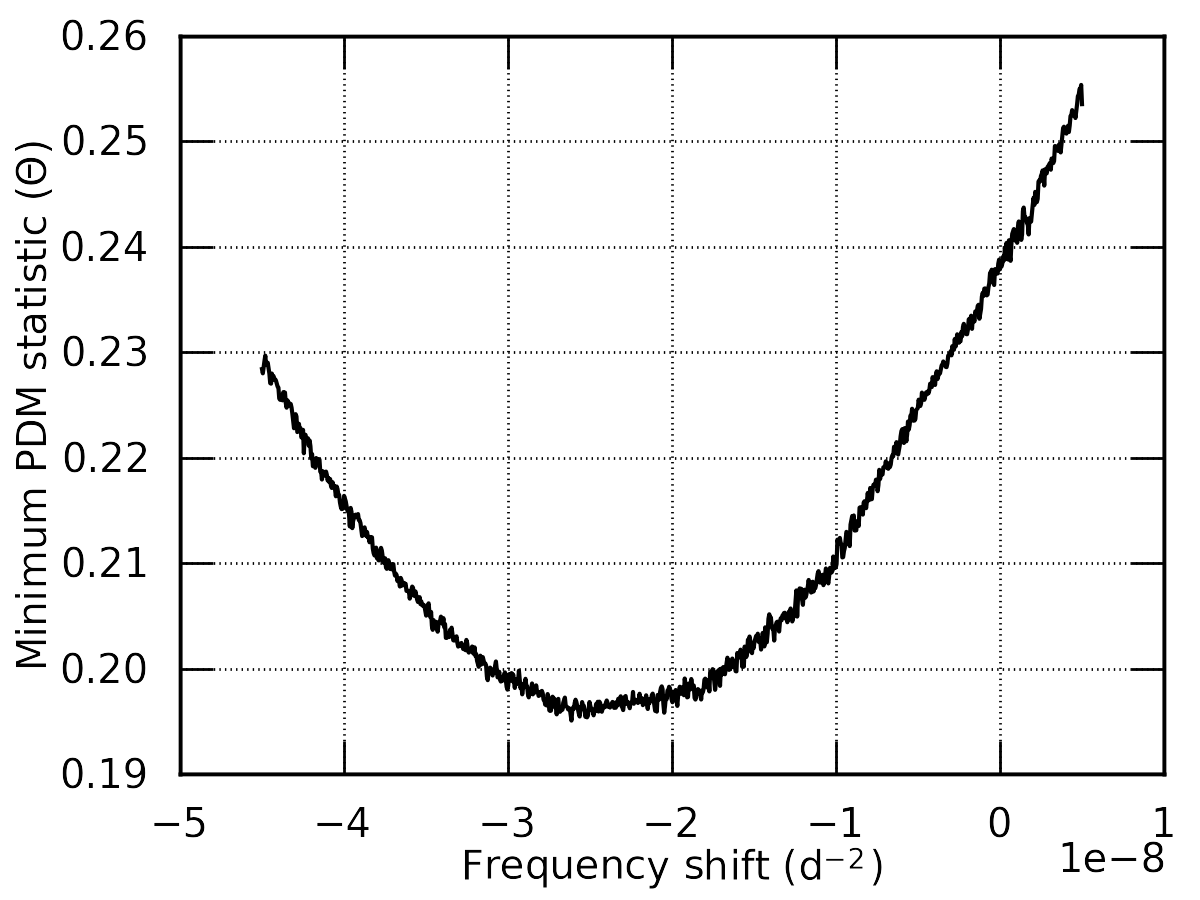}

\includegraphics[width=0.49\columnwidth]{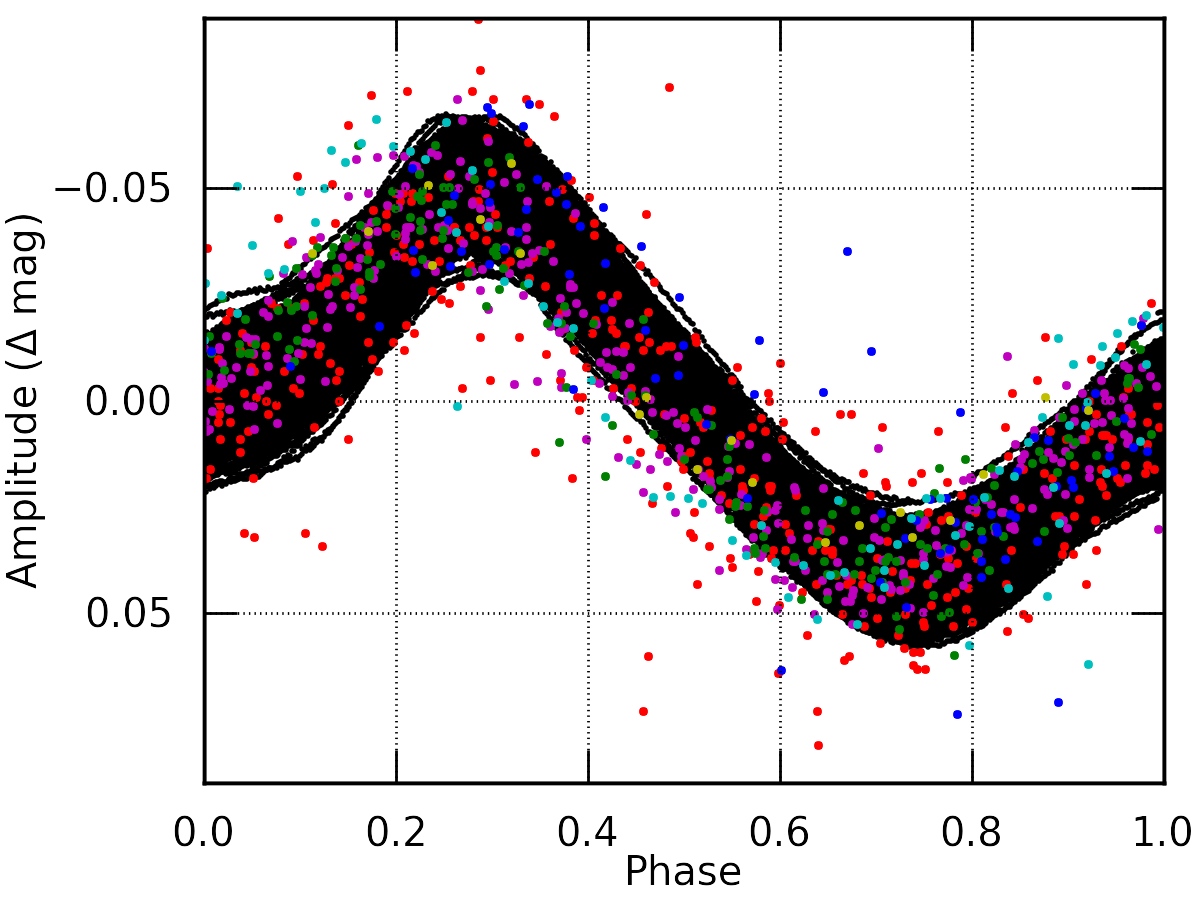}
\includegraphics[width=0.49\columnwidth]{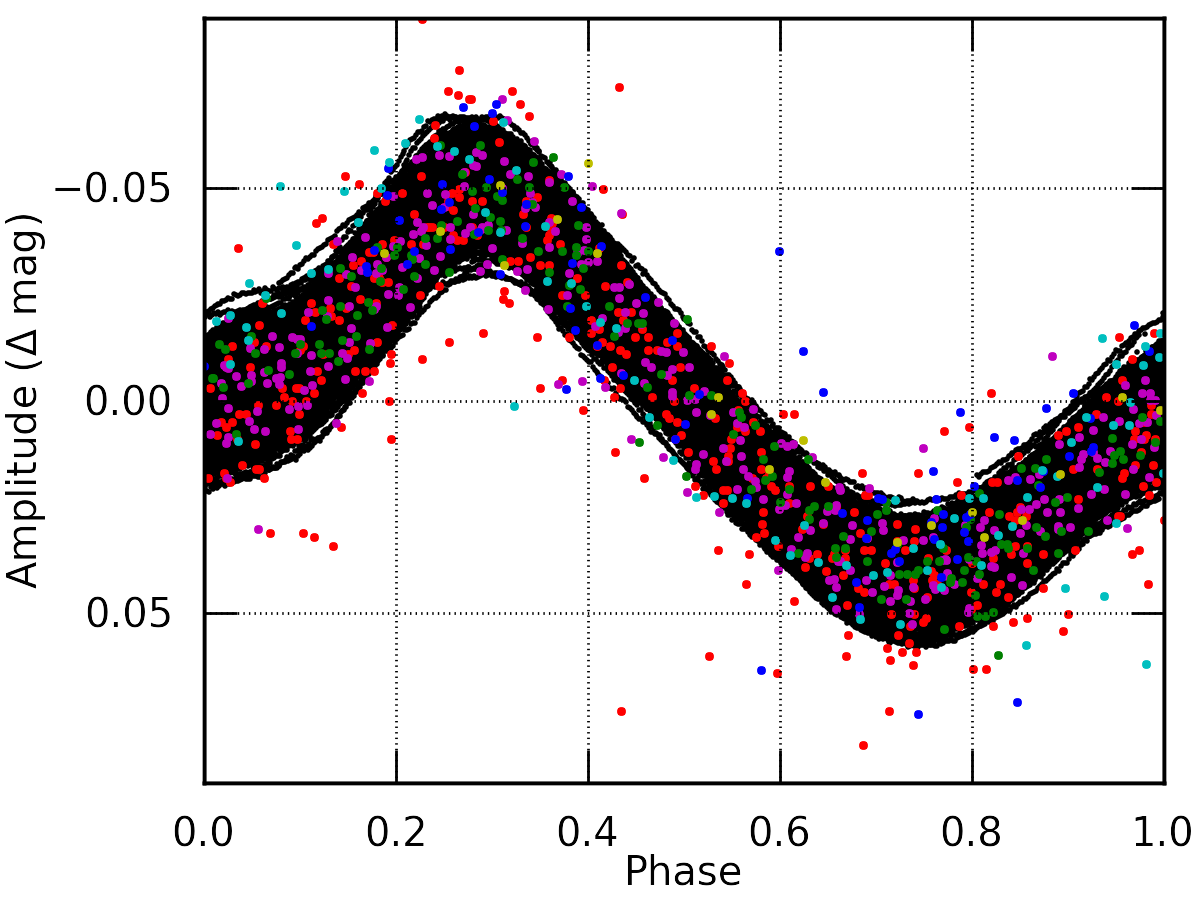}

\caption{\emph{Upper left}: Frequency determinations of ground and space based observations
(same colour coding as in Fig.\,\ref{fig:hd180642:alldata}). Vertical bars denote $1\sigma$ error in frequency, corrected for correlation. Horizontal bars denote time span for the frequency determination,
circles denote midpoint of observations. The offset on the frequency from the Konkoly data is a result of a bad window function.  \emph{Upper right}: minimum PDM test statistic as a function of trial frequency shift. \emph{Lower left}: phase diagram with $D=0$\,d$^{-2}$. \emph{Lower right}: phase diagram with $D-2.6\times 10^{-8}$\,d$^{-2}$.}\label{fig:hd180642:freqsummary}
\end{figure}

To quantify the decrease, I assume a simple linear decrease and apply an adapted phase dispersion minimization (PDM) method \citep{stellingwerf1978,cuypers1986} to the combined datasets, omitting the CoRoT dataset. The latter is only used to check the findings, since the high cadence and relatively short time completely dominate the frequency search procedures. I test for 1000 trial frequency shifts $D$ between $-4.5\times 10^{-8}$\,d$^{-2}$ and $0.5\times 10^{-8}$\,d$^{-2}$ in steps of $5\times 10^{-11}$\,d$^{-2}$ and calculate at what frequency the PDM test statistic ($\Theta$) reaches its minimum (Fig.\,\ref{fig:hd180642:freqsummary}, top right panel). A clear minimum is found around $D_1= (-2.3\pm0.2)\times 10^{-8}$\,d$^{-2}$ (a period increase of 2.4 seconds per century), which corresponds to a reduced scatter in the phase diagram compared to $D_1=0$\,d$^{-2}$ (Fig.\,\ref{fig:hd180642:freqsummary}, bottom panels). The error on the parameter is determined via the covariance matrix obtained from the nonlinear optimization routine, and are confirmed with those obtained from an F-test applied on the $\chi^2$ statistic in an interval around the parameters.

Excluding the Hipparcos measurements decreases the time span of the observations but yields consistent results ($D_2=(-3.0\pm0.5)\times10^{-8}$\,d$^{-2}$). Including the entire CoRoT dataset also yields consistent results within a 3$\sigma$ confidence interval ($D_3=(-2.61\pm0.06)\times10^{-8}$\,d$^{-2}$). Regardless of the included datasets, there is an increase in variance reduction and a decrease in AIC and BIC criterion \citep[for a definition, see, e.g., ][]{degroote2009b_mnras} when allowing for a frequency shift, confirming that a linearly decreasing frequency gives a better description of the data.
The sparseness of the data sets, however, make it difficult to assess if the frequency shift is linear, quadratic, periodic or following yet a different pattern. The simple conclusion is that assuming a variable frequency, the data is significantly better described than without it.

Possible causes for artificial frequency shifts between two datasets well separated in time, can be different time conventions \citep[e.g,][, Chapter 4]{aerts2009book}. However, the results are insensitive to small errors in the calculations of the time points. Deviations as large as 10\,s translate to a shift in the phase diagram of only 0.06\%, which has no influence on the results. Also, the results obtained from the different datasets from Hipparcos, CoRoT and ASAS separately, show a frequency decrease which is independent from any mismatch between different time coordinate systems (e.g., Julian Date or Modified Julian date). A two-sample $t$-test shows that the probability that the frequencies in the Hipparcos and ASAS data are equal is below $0.1\%$. Finally, the aliasing problem which is present in the Konkoly data set, is not present in the Hipparcos and ASAS data sets, eliminating the possibility that aliasing problems are at the basis of the frequency differences. Thus, the observed frequency shift must be intrinsic to the target.

A first physical explanation for period changes is light-time effects due to binarity. Light-time effects predict a radial velocity of
 \[v_\mathrm{rad} = c\frac{\Delta f}{f},\]
 translating to a radial velocity change of $\sim 10$\,km\,s$^{-1}$ over the 20 years of photometric observations of HD\,180642. The datasets currently available are not sufficient to reveal possible modulations. However, there are no signs of binarity in the high signal-to-noise, long term spectroscopic campaign done by \citet{briquet2009_mnras}, putting tight constraints on the luminosity of a hypothetical secondary object.

I confront the period increase found in the combined datasets with expected values from a grid of evolutionary models \citep[see also ][]{aerts2011} calculated with the CLES \citep{cles} code, where the adiabatic approximation is used to calculate the frequencies with the LOSC code \citep{scuflaire2008b} . The evolutionary frequency shift was calculated by extracting all $\ell=0$ modes from the grid, and taking the numerical 3-point Lagrangian derivative along the evolutionary track. As the star slowly expands during the main sequence, its mean density decreases together with the frequencies of the radial modes. The observed frequency decrease is at least 10 times larger than models predict, except for stars in the hydrogen shell burning phase. Previous seismic modelling excludes these stages, and given the short time scale of the hydrogen shell burning stage, it is unlikely to observe a star during this stage.

After arguing against an instrumental, evolutionary and external explanation for the frequency shift, I propose another possibility: nonlinear effects. Although nonlinear effects are not typically expected to produce consistent long term frequency variations, slight changes in phase timing can mimic long term trends when the observations are divided in distinct groups containing large gaps. In the following sections, I look for more signs of nonlinearity, and argue that this hypothesis can provide an alternative explanation also for the power excess observed in the frequency spectrum. 

\subsection{Power excess}

A standard frequency analysis of the CoRoT light curve with iterative prewhitening results in thousands of formally significant peaks. Such a model is not very meaningful in view of the limited frequency resolution. In contrast to the fairly constant-amplitude background excess in the frequency spectrum observed in HD\,50844 \citep{poretti2009_mnras}, the spectrum of HD\,180642 is highly structured, even after prewhitening 500 frequencies. \citet{kallinger2010} suggested granulation to be the cause for the background power observed in two A-type stars. In the case of HD\,180642, that model does not give a satisfactory fit to the structures. \citet{smolec2007} suggested that amplitude saturation could play an important role in exciting hundreds of resonant modes in $\beta$\,Cephei stars, perhaps even into the gravity mode regime. The number of observed frequencies in HD\,180642 is, however, much larger, and the range of frequencies is large compared to the range predicted by excitation calculations. On the other hand, \citep{buchler1996} interpreted the multiperiodicity in the RV Tauri star R\,Scuti as low-dimensional chaos, and not as actually excited modes. 

In the following, I investigate if the nonlinear behaviour of the dominant radial mode could play a role in the power excess in the frequency spectrum. Confirmation of such nonlinearity could also possibly explain the observed shift in frequency as due to slight irregularities in the dominant frequency: if the dominant mode is suffering from chaos, it is expected that not every pulsation cycle repeats itself exactly, which can mimick long-term frequency or phase shifts in grouped observations such as the archival data set at hand.

It is known that even simple oscillators, e.g. those that are driven and damped, can show non-repetitive behaviour, as long as the amplitude of oscillator is in a particular range. \citet{tanaka1988} proposed and investigated a simple 1D model for nonlinear pulsations, which was used for simulations in \citet{votruba2009}. There, it was shown that chaotic oscillations can introduce spurious periods in the periodogram. Their nonadiabatic oscillator is described by
\begin{equation}\frac{d^2x(t)}{dt^2} -\mu\frac{dx(t)}{dt} - \omega^2\,x(t) = z(t),\label{eq:nonlin_nonad_osc1}\end{equation}
where $x(t)$ describes the surface displacement, $\mu$ is the driving factor, $\omega$ is the frequency and $z(t)$ the driving force. \citet{tanaka1988} set
\begin{equation}\frac{dz(t)}{dt} = -\beta\frac{dx(t)}{dt} - p\,z(t) - q\frac{dx(t)}{dt} + s\,z(t)\frac{dx(t)}{dt}.\label{eq:nonlin_nonad_osc}\end{equation}
The parameter $p$ mimics the energy lost during a pulsation cycle, e.g., via radiation, while the nonlinear term
preceded by the parameter $s$ introduces a phase shift between the cooling and heating. The system can be solved using a a standard iterative Runge-Kutta method. As an example I set $\omega^2=0.5$, $\mu=-0.5$ (driving), $p=4.0$ (radiative cooling), and $s=1.0$ (phase shift between cooling and heating) by default, and only change how the material responds to volume changes via the parameter $\beta$. High $\beta$ values result in a strong drop of driving force when the material expands, in which case the amplitude of the oscillation should be low and thus the linear approximation is valid. Indeed, a value of $\beta=2$ gives a monoperiodic oscillator with a sinusoidal phase shape. On the other hand, a value of $\beta=0.7$ drives the amplitudes to much higher values into the nonlinear regime, resulting in a distorted phase shape.

To mimic the behaviour of HD\,180642, I start from a set of parameters used in \citet{tanaka1988}: $a=4, b=5, \mu=-0.5, q=-0.7$, and $s=5$. Changing the value of $p$ between $-0.04$ and $-0.08$ creates a Feigenbaum diagram, that allows to carefully pick a value that restricts the size of the attractor in phase space, as is observed for HD\,180642. I choose a value $p=-0.726$, emphasizing that the setup of the simulation is only chosen to qualitatively study the influence of nonlinear behaviour on the frequency spectrum. The aim is not to quantitatively model the dominant frequency of HD\,180642 using Eq. \ref{eq:nonlin_nonad_osc}.

\begin{figure}
\centering
\includegraphics[width=\columnwidth]{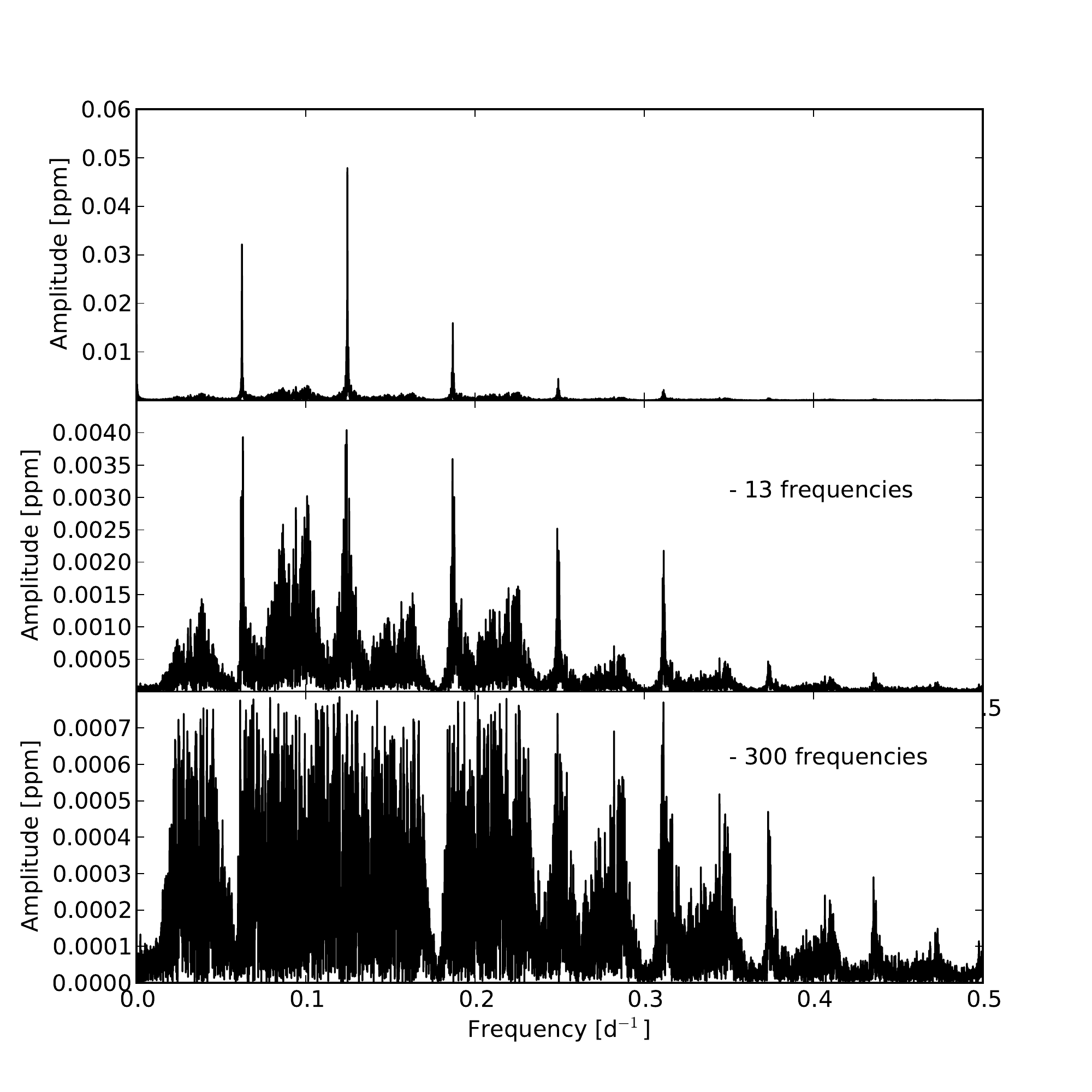}
\caption{Original frequency spectrum of a simulated nonlinear oscillator containing chaos (top panel), after removing 13 frequencies (middle panel) and after removing 300 frequencies.}\label{fig:hd180642:resper}
\end{figure}

I generate a time series of 175\,200 points, add Gaussian noise with a standard deviation similar to the CoRoT data, and apply standard iterative prewhitening tools as it was done for the stellar data in \citep{degroote2009b_mnras} to the simulations. In Fig.\,\ref{fig:hd180642:resper}, the original frequency spectrum is shown (top panel), where the harmonics and half-frequency (from the period-doubling phenomenon) are clearly visible. Some power excess is visible over the entire range of harmonics. Iteratively removing 13 frequencies makes the power excess at low amplitudes better visible. In a similar way as was found for the CoRoT light curve of HD\,180642, the number of frequencies from an iterative prewhitening analysis amounts to several hundreds to thousands. In contrast to red noise components or granulation signal, the background amplitude spectrum does not appear as a plateau, but rather as a group of dense peaks.

\begin{figure}
\centering
\includegraphics[width=\columnwidth]{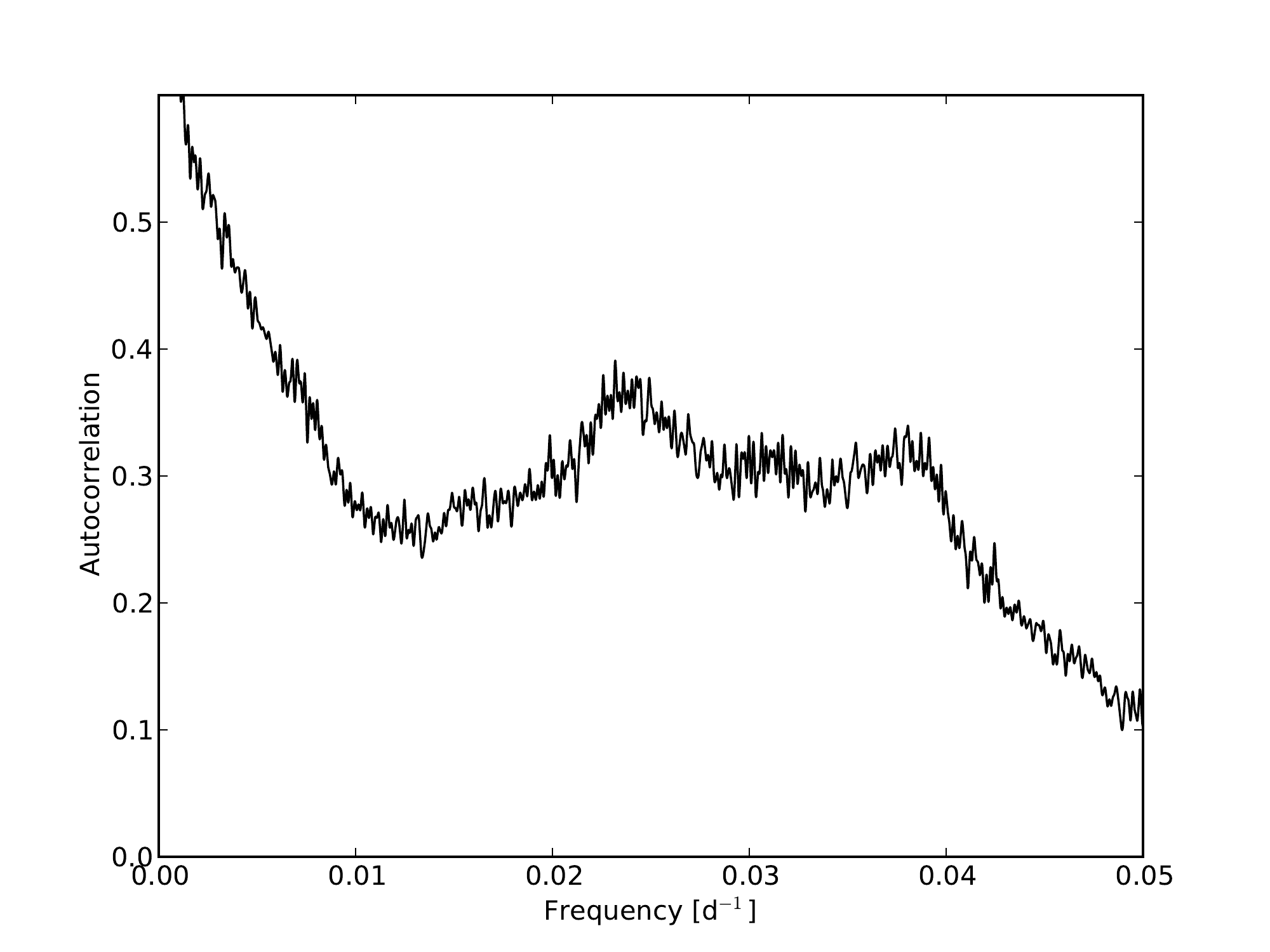}
\caption{Autocorrelation of residual frequency spectrum.}\label{fig:hd180642:ac}
\end{figure}

The previous exploration shows that the frequency spectrum is contaminated by a huge number of high signal-to-noise ratio peaks, which are not connected to eigenfrequencies of the system (there is only one in the simulated system). An autocorrelation (Fig.\,\ref{fig:hd180642:ac}) of the frequency spectrum reveals structure, which is similar to what is expected from stochastic pulsations \citep[e.g., ][]{carrier2003}, and remarkably similar to the one shown in \citet[][see the supplementary material]{belkacem2009}. In the next Section, I attempt at quantifying to what extent HD\,180642 exhibits nonlinear behaviour.

\subsection{Nonlinear analysis}
Chaotic signals can be recognised by their characteristic phase space behaviour. While monoperiodic oscillators have only one orbit in phase-space (i.e., each cycle repeats exactly), chaotic attractors appear in the case of the chaotic oscillators.

\begin{figure}
 \centering
 \includegraphics[width=\columnwidth]{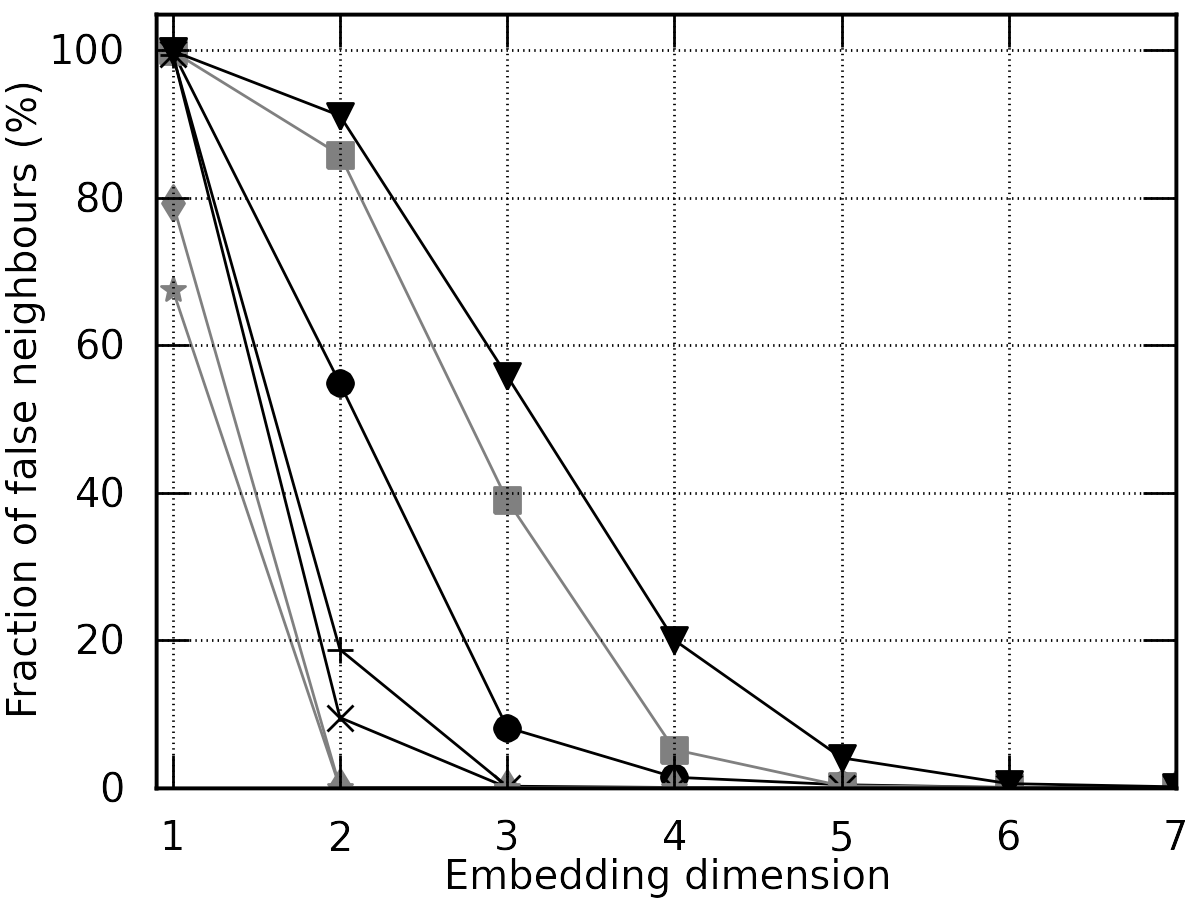}
\caption{Determination of embedding dimension using the false neighbour method ($*$=nonlinear nonadiabatic oscillator (NNO) with $\beta=2$, diamond: NNO with $\beta=0.7$, $\times=$NNO with $\beta=1$, plus: NNO with $\beta=0.8$, bullet: Lorenz\,II \citep{lorenz1984,kennel1992}, black square: multiperiodic signal, black triangle: pure noise). Since multiperiodic signals also require a high embedding dimension, the false neighbour method cannot distinguish these types of signals from true chaotic signals.}
 \label{fig:methodology:false_neighbours}
\end{figure}

When the dynamics of the system is known, the phase-space construction is straightforward and chaotic behaviour can be readily determined. In the case of observed time series, however, a different approach has to be followed. Deterministic chaotic signals can be characterised by the largest Lyaponuv exponent ($\lambda_1$), which is the average exponential rate of divergence (or convergence) between two nearby points in phase space \citep{wolf1985},
\[d(t) = C\exp(\lambda_1 t),\]
where $C$ is a normalisation constant. Lyapunov exponents thus deliver a measure of how quickly two neighbouring points in phase space separate. Strictly periodic signals have $\lambda_1=0$, because the orbits of any two points are the same. When $\lambda_1<0$, the orbits converge (e.g. in the case of a damped oscillator), and if $\lambda_1>0$, the orbits are unstable and chaotic. The basic steps to determine the maximum Lyapunov exponent thus consist of reconstructing phase-space and measuring the rate of divergence. I follow the false neighbour method approach of \citet{rosenstein1993}, which is fast, easy to implement, and reliable also in the case of small datasets. Extensive numerical tests on systems of different chaotic degrees (i.e systems for which different embedding dimensions are required) as well as light curves with a highly multiperiodic signal, have shown that also the latter require high embedding dimensions, rendering nonlinear time series analysis techniques on light curves containing highly multiperiodic signal less useful. In Fig.\,\ref{fig:methodology:false_neighbours}, the influence of the choice of embedding dimension is shown on the number of false neighbours, as is the influence of noise. For monoperiodic, the embedding dimension is very low, since the number of false neighbours drops quickly to zero. In highly nonlinear systems, the number of false neighbours drops only significantly below 100\% at high values of the embedding dimensions. However, we see the same behaviour for highly multiperiodic (but linear) systems.

I conclude that, although it is very promising to separate monoperiodic signals and noise from chaotic time series, this is not the case for highly multiperiodic signals. Based only on the method of false neighbours and in the presence of noise, the same behaviour is found for the Lorenz\,II equations \citep{lorenz1984,kennel1992} and the multiperiodic signal. The limiting factor for the separability of multiperiodicity and chaos will thus be the noise level and frequency resolution (or the number of cycles observed).

Choosing embedding dimensions between $d_e=2$ and $d_e=7$ and an embedding delay of $\tau\approx0.02$\,d \citep[the value where the autocorrelation function drops below $1-1/e$, ][]{rosenstein1993}, there is a high exponential separation rate, which reaches saturation level during the short time span of one orbit. This result is not sensitive to the number of modes we prewhiten beforehand to minimize the influence of multiperiodicity, nor to the choice of the embedding delay $\tau$. I conclude that the frequency spectrum of HD\,180642 is dominated by the signal from actual pulsation modes, so that nonlinear time series analysis techniques have difficulties detecting chaotic behaviour unambiguously in the CoRoT light curve of HD\,180642.

\section{Conclusion}

In this paper, I proposed the presence of nonlinear behaviour as an alternative explanation for much of the low-amplitude power excess at low and high frequencies in the CoRoT light curve of HD\,180642. The first explanation by \citet{belkacem2009} invoked the presence of solar-like oscillations, but failed to explain the lower frequency power excess. The second explanation was given by \citet{aerts2011} via nonlinear resonance, but requires the presence of a dense spectrum of eigenmodes, containing many thousands of frequencies.

It was not possible to unambiguously determine the origin of the power excess, but from the point of view of model simplicity, I argue in favour of chaos, since it explains much of the observed behaviour without invoking new physical processes. At least four anomalies in the observations of HD\,180642, compared to the expected relative scarce and structured frequency spectrum delivered by linear theory, are explained by the chaos hypothesis. First, it explains the observed period change, which is not compatible with current stellar evolution models, as a natural side effect of nonlinear behaviour. This is perhaps also applicable to the observed smooth period changes observed in RR\,Lyrae stars by \citet{kolenberg2010_mnras} or the scatter observed by \citet{szabo2011_mnras} in Cepheids, or by \citet{zijlstra2004} in Mira variables. All of these variables show large amplitude variations, driven far into the nonlinear regime. Second, it explains at least qualitatively the signal in the autocorrelation. In contrast to red noise components or granulation, chaos can introduce structure in the form of dense forests of peaks in the power spectrum, introducing spurious peaks in the autocorrelation diagram. Third, it explains the wealth of peaks in a wide frequency range, that are not necessarily eigenfrequencies. Finally, if all the frequencies in the Fourier spectrum are \emph{true} frequencies excited in the star by any mechanism, the amplitude saturation calculations of \citet{smolec2007} would most likely predict a lower saturation amplitude in the case of HD\,180642. Their results show that the saturation amplitude roughly scales as $n^{-0.5}$ with $n$ the number of excited frequencies, and decreases even further with lower metal abundances. The observed visual amplitude of HD\,180642 is at the high end  of their calculations with collective saturation (their Fig.\,7), while the number of observed frequencies in the pressure and gravity mode regime is at least an order of magnitude larger and the observed metal abundance is significantly sub-solar \citep{briquet2009_mnras}. The presence of \emph{artificial} frequencies in the light curve arising from the chaos, would strongly decrease the \emph{true} number of modes taking part in the saturation effect, making the CoRoT observations more in line with the predictions of \citet{smolec2007}.

A possible way forward from the observational point of view would be to continue long term monitoring of the dominant mode of HD\,180642, which would allow us to check if the evolution of the frequency follows a consistent trend or is rather of chaotic nature. Most likely, the GAIA satellite will observe this star at least occasionally for an extended period of time, which will, dependent on the launch date, add at least another ten years to the time span. The GAIA satellite will then also observe other high-amplitude $\beta$\,Cep stars, which would allow us to check if these period changes are common in $\beta$\,Cephei stars, even for low amplitude modes. If no correlation with amplitude can be found, the idea of short-term evolutionary fluctuations (`evolutionary weather') should be favoured. From theory, it could be worthwile to apply specifically tuned nonlinear hydrodynamical models as presented by \citet{smolec2007}, to reproduce the observed amplitude of the dominant mode.

Still, the nonlinear timeseries analysis showed that the nonlinear behaviour, if present, is definitely not strong in the CoRoT data. The amplitude of the power excess is much lower than the amplitudes of the strongest modes (especially the dominant radial mode), and most of the high signal-to-noise peaks are due to additional pulsation modes. This is also confirmed by the successful modelling by \citet{aerts2011}. We cannot, however, confidently identify which peaks of the low amplitude signal is due to pulsations, and which are more likely to arise from the nonlinear behaviour.

\subsection*{Acknowledgments}

Pieter Degroote is a postdoctoral fellow of the Fund for Scientific Research, Flanders (FWO).

\bibliographystyle{mn}

\label{lastpage}

\end{document}

%% file: ref_abbrevs.tex
\def\aap{A\&A}%